\documentclass[letterpaper,compsoc,twoside]{IEEEtran}
\usepackage{fixltx2e} 
\usepackage{cmap} 
\usepackage{ifthen}
\usepackage[T1]{fontenc}
\usepackage[utf8]{inputenc}
\usepackage{amsmath}

\usepackage[font={small,it},labelfont=bf]{caption}
\usepackage{float}

\usepackage{scipy}
\makeatletter
\def\PY@reset{\let\PY@it=\relax \let\PY@bf=\relax%
    \let\PY@ul=\relax \let\PY@tc=\relax%
    \let\PY@bc=\relax \let\PY@ff=\relax}
\def\PY@tok#1{\csname PY@tok@#1\endcsname}
\def\PY@toks#1+{\ifx\relax#1\empty\else%
    \PY@tok{#1}\expandafter\PY@toks\fi}
\def\PY@do#1{\PY@bc{\PY@tc{\PY@ul{%
    \PY@it{\PY@bf{\PY@ff{#1}}}}}}}
\def\PY#1#2{\PY@reset\PY@toks#1+\relax+\PY@do{#2}}

\expandafter\def\csname PY@tok@gd\endcsname{\def\PY@tc##1{\textcolor[rgb]{0.63,0.00,0.00}{##1}}}
\expandafter\def\csname PY@tok@gu\endcsname{\let\PY@bf=\textbf\def\PY@tc##1{\textcolor[rgb]{0.50,0.00,0.50}{##1}}}
\expandafter\def\csname PY@tok@gt\endcsname{\def\PY@tc##1{\textcolor[rgb]{0.00,0.27,0.87}{##1}}}
\expandafter\def\csname PY@tok@gs\endcsname{\let\PY@bf=\textbf}
\expandafter\def\csname PY@tok@gr\endcsname{\def\PY@tc##1{\textcolor[rgb]{1.00,0.00,0.00}{##1}}}
\expandafter\def\csname PY@tok@cm\endcsname{\let\PY@it=\textit\def\PY@tc##1{\textcolor[rgb]{0.25,0.50,0.56}{##1}}}
\expandafter\def\csname PY@tok@vg\endcsname{\def\PY@tc##1{\textcolor[rgb]{0.73,0.38,0.84}{##1}}}
\expandafter\def\csname PY@tok@m\endcsname{\def\PY@tc##1{\textcolor[rgb]{0.13,0.50,0.31}{##1}}}
\expandafter\def\csname PY@tok@mh\endcsname{\def\PY@tc##1{\textcolor[rgb]{0.13,0.50,0.31}{##1}}}
\expandafter\def\csname PY@tok@cs\endcsname{\def\PY@tc##1{\textcolor[rgb]{0.25,0.50,0.56}{##1}}\def\PY@bc##1{\setlength{\fboxsep}{0pt}\colorbox[rgb]{1.00,0.94,0.94}{\strut ##1}}}
\expandafter\def\csname PY@tok@ge\endcsname{\let\PY@it=\textit}
\expandafter\def\csname PY@tok@vc\endcsname{\def\PY@tc##1{\textcolor[rgb]{0.73,0.38,0.84}{##1}}}
\expandafter\def\csname PY@tok@il\endcsname{\def\PY@tc##1{\textcolor[rgb]{0.13,0.50,0.31}{##1}}}
\expandafter\def\csname PY@tok@go\endcsname{\def\PY@tc##1{\textcolor[rgb]{0.20,0.20,0.20}{##1}}}
\expandafter\def\csname PY@tok@cp\endcsname{\def\PY@tc##1{\textcolor[rgb]{0.00,0.44,0.13}{##1}}}
\expandafter\def\csname PY@tok@gi\endcsname{\def\PY@tc##1{\textcolor[rgb]{0.00,0.63,0.00}{##1}}}
\expandafter\def\csname PY@tok@gh\endcsname{\let\PY@bf=\textbf\def\PY@tc##1{\textcolor[rgb]{0.00,0.00,0.50}{##1}}}
\expandafter\def\csname PY@tok@ni\endcsname{\let\PY@bf=\textbf\def\PY@tc##1{\textcolor[rgb]{0.84,0.33,0.22}{##1}}}
\expandafter\def\csname PY@tok@nl\endcsname{\let\PY@bf=\textbf\def\PY@tc##1{\textcolor[rgb]{0.00,0.13,0.44}{##1}}}
\expandafter\def\csname PY@tok@nn\endcsname{\let\PY@bf=\textbf\def\PY@tc##1{\textcolor[rgb]{0.05,0.52,0.71}{##1}}}
\expandafter\def\csname PY@tok@no\endcsname{\def\PY@tc##1{\textcolor[rgb]{0.38,0.68,0.84}{##1}}}
\expandafter\def\csname PY@tok@na\endcsname{\def\PY@tc##1{\textcolor[rgb]{0.25,0.44,0.63}{##1}}}
\expandafter\def\csname PY@tok@nb\endcsname{\def\PY@tc##1{\textcolor[rgb]{0.00,0.44,0.13}{##1}}}
\expandafter\def\csname PY@tok@nc\endcsname{\let\PY@bf=\textbf\def\PY@tc##1{\textcolor[rgb]{0.05,0.52,0.71}{##1}}}
\expandafter\def\csname PY@tok@nd\endcsname{\let\PY@bf=\textbf\def\PY@tc##1{\textcolor[rgb]{0.33,0.33,0.33}{##1}}}
\expandafter\def\csname PY@tok@ne\endcsname{\def\PY@tc##1{\textcolor[rgb]{0.00,0.44,0.13}{##1}}}
\expandafter\def\csname PY@tok@nf\endcsname{\def\PY@tc##1{\textcolor[rgb]{0.02,0.16,0.49}{##1}}}
\expandafter\def\csname PY@tok@si\endcsname{\let\PY@it=\textit\def\PY@tc##1{\textcolor[rgb]{0.44,0.63,0.82}{##1}}}
\expandafter\def\csname PY@tok@s2\endcsname{\def\PY@tc##1{\textcolor[rgb]{0.25,0.44,0.63}{##1}}}
\expandafter\def\csname PY@tok@vi\endcsname{\def\PY@tc##1{\textcolor[rgb]{0.73,0.38,0.84}{##1}}}
\expandafter\def\csname PY@tok@nt\endcsname{\let\PY@bf=\textbf\def\PY@tc##1{\textcolor[rgb]{0.02,0.16,0.45}{##1}}}
\expandafter\def\csname PY@tok@nv\endcsname{\def\PY@tc##1{\textcolor[rgb]{0.73,0.38,0.84}{##1}}}
\expandafter\def\csname PY@tok@s1\endcsname{\def\PY@tc##1{\textcolor[rgb]{0.25,0.44,0.63}{##1}}}
\expandafter\def\csname PY@tok@gp\endcsname{\let\PY@bf=\textbf\def\PY@tc##1{\textcolor[rgb]{0.78,0.36,0.04}{##1}}}
\expandafter\def\csname PY@tok@sh\endcsname{\def\PY@tc##1{\textcolor[rgb]{0.25,0.44,0.63}{##1}}}
\expandafter\def\csname PY@tok@ow\endcsname{\let\PY@bf=\textbf\def\PY@tc##1{\textcolor[rgb]{0.00,0.44,0.13}{##1}}}
\expandafter\def\csname PY@tok@sx\endcsname{\def\PY@tc##1{\textcolor[rgb]{0.78,0.36,0.04}{##1}}}
\expandafter\def\csname PY@tok@bp\endcsname{\def\PY@tc##1{\textcolor[rgb]{0.00,0.44,0.13}{##1}}}
\expandafter\def\csname PY@tok@c1\endcsname{\let\PY@it=\textit\def\PY@tc##1{\textcolor[rgb]{0.25,0.50,0.56}{##1}}}
\expandafter\def\csname PY@tok@kc\endcsname{\let\PY@bf=\textbf\def\PY@tc##1{\textcolor[rgb]{0.00,0.44,0.13}{##1}}}
\expandafter\def\csname PY@tok@c\endcsname{\let\PY@it=\textit\def\PY@tc##1{\textcolor[rgb]{0.25,0.50,0.56}{##1}}}
\expandafter\def\csname PY@tok@mf\endcsname{\def\PY@tc##1{\textcolor[rgb]{0.13,0.50,0.31}{##1}}}
\expandafter\def\csname PY@tok@err\endcsname{\def\PY@bc##1{\setlength{\fboxsep}{0pt}\fcolorbox[rgb]{1.00,0.00,0.00}{1,1,1}{\strut ##1}}}
\expandafter\def\csname PY@tok@kd\endcsname{\let\PY@bf=\textbf\def\PY@tc##1{\textcolor[rgb]{0.00,0.44,0.13}{##1}}}
\expandafter\def\csname PY@tok@ss\endcsname{\def\PY@tc##1{\textcolor[rgb]{0.32,0.47,0.09}{##1}}}
\expandafter\def\csname PY@tok@sr\endcsname{\def\PY@tc##1{\textcolor[rgb]{0.14,0.33,0.53}{##1}}}
\expandafter\def\csname PY@tok@mo\endcsname{\def\PY@tc##1{\textcolor[rgb]{0.13,0.50,0.31}{##1}}}
\expandafter\def\csname PY@tok@mi\endcsname{\def\PY@tc##1{\textcolor[rgb]{0.13,0.50,0.31}{##1}}}
\expandafter\def\csname PY@tok@kn\endcsname{\let\PY@bf=\textbf\def\PY@tc##1{\textcolor[rgb]{0.00,0.44,0.13}{##1}}}
\expandafter\def\csname PY@tok@o\endcsname{\def\PY@tc##1{\textcolor[rgb]{0.40,0.40,0.40}{##1}}}
\expandafter\def\csname PY@tok@kr\endcsname{\let\PY@bf=\textbf\def\PY@tc##1{\textcolor[rgb]{0.00,0.44,0.13}{##1}}}
\expandafter\def\csname PY@tok@s\endcsname{\def\PY@tc##1{\textcolor[rgb]{0.25,0.44,0.63}{##1}}}
\expandafter\def\csname PY@tok@kp\endcsname{\def\PY@tc##1{\textcolor[rgb]{0.00,0.44,0.13}{##1}}}
\expandafter\def\csname PY@tok@w\endcsname{\def\PY@tc##1{\textcolor[rgb]{0.73,0.73,0.73}{##1}}}
\expandafter\def\csname PY@tok@kt\endcsname{\def\PY@tc##1{\textcolor[rgb]{0.56,0.13,0.00}{##1}}}
\expandafter\def\csname PY@tok@sc\endcsname{\def\PY@tc##1{\textcolor[rgb]{0.25,0.44,0.63}{##1}}}
\expandafter\def\csname PY@tok@sb\endcsname{\def\PY@tc##1{\textcolor[rgb]{0.25,0.44,0.63}{##1}}}
\expandafter\def\csname PY@tok@k\endcsname{\let\PY@bf=\textbf\def\PY@tc##1{\textcolor[rgb]{0.00,0.44,0.13}{##1}}}
\expandafter\def\csname PY@tok@se\endcsname{\let\PY@bf=\textbf\def\PY@tc##1{\textcolor[rgb]{0.25,0.44,0.63}{##1}}}
\expandafter\def\csname PY@tok@sd\endcsname{\let\PY@it=\textit\def\PY@tc##1{\textcolor[rgb]{0.25,0.44,0.63}{##1}}}

\def\PYZbs{\char`\\}
\def\PYZus{\char`\_}
\def\PYZob{\char`\{}
\def\PYZcb{\char`\}}

\def\PYZam{\char`\&}
\def\PYZlt{\char`\<}
\def\PYZgt{\char`\>}
\def\PYZsh{\char`\#}

\def\PYZhy{\char`\-}
\def\PYZsq{\char`\'}
\def\PYZdq{\char`\"}
\def\PYZti{\char`\~}

\makeatother

\providecommand*{\DUfootnotemark}[3]{%
  \raisebox{1em}{\hypertarget{#1}{}}%
  \hyperlink{#2}{\textsuperscript{#3}}%
}
\providecommand{\DUfootnotetext}[4]{%
  \begingroup%
  \renewcommand{\thefootnote}{%
    \protect\raisebox{1em}{\protect\hypertarget{#1}{}}%
    \protect\hyperlink{#2}{#3}}%
  \footnotetext{#4}%
  \endgroup%
}

\providecommand*{\DUrole}[2]{%
  \ifcsname DUrole#1\endcsname%
    \csname DUrole#1\endcsname{#2}%
  \else
    \ifcsname docutilsrole#1\endcsname%
      \csname docutilsrole#1\endcsname{#2}%
    \else%
      #2%
    \fi%
  \fi%
}

\ifthenelse{\isundefined{\hypersetup}}{
  \usepackage[colorlinks=true,linkcolor=blue,urlcolor=blue]{hyperref}
  \urlstyle{same} 
}{}

\begin{document}
\newcounter{footnotecounter}\title{Frequentism and Bayesianism: A Python-driven Primer}\author{Jake VanderPlas$^{\setcounter{footnotecounter}{1}\fnsymbol{footnotecounter}\setcounter{footnotecounter}{2}\fnsymbol{footnotecounter}}$%
          \setcounter{footnotecounter}{1}\thanks{\fnsymbol{footnotecounter} %
          Corresponding author: \protect\href{mailto:jakevdp@cs.washington.edu}{jakevdp@cs.washington.edu}}\setcounter{footnotecounter}{2}\thanks{\fnsymbol{footnotecounter} eScience Institute, University of Washington}\thanks{%

          \noindent%
          Copyright\,\copyright\,2014 Jake VanderPlas. This is an open-access article distributed under the terms of the Creative Commons Attribution License, which permits unrestricted use, distribution, and reproduction in any medium, provided the original author and source are credited.%
        }\\
}\maketitle
          \renewcommand{\leftmark}{PROC. OF THE 13th PYTHON IN SCIENCE CONF. (SCIPY 2014)}
          \renewcommand{\rightmark}{FREQUENTISM AND BAYESIANISM: A PYTHON-DRIVEN PRIMER}

\InputIfFileExists{page_numbers.tex}{}{}
\newcommand*{\docutilsroleref}{\ref}
\newcommand*{\docutilsrolelabel}{\label}
\begin{abstract}This paper presents a brief, semi-technical comparison of the essential features of the frequentist and Bayesian approaches to statistical inference, with several illustrative examples implemented in Python. The differences between frequentism and Bayesianism fundamentally stem from differing definitions of probability, a philosophical divide which leads to distinct approaches to the solution of statistical problems as well as contrasting ways of asking and answering questions about unknown parameters. After an example-driven discussion of these differences, we briefly compare several leading Python statistical packages which implement frequentist inference using classical methods and Bayesian inference using Markov Chain Monte Carlo.\DUfootnotemark{id1}{blog}{1}\end{abstract}%
\DUfootnotetext{blog}{id1}{1}{\phantomsection\label{blog}
This paper draws heavily from content originally published in a series of posts on the author's blog, \href{http://jakevdp.github.io/}{Pythonic Perambulations} \cite{VanderPlas2014}.}

\subsection{Introduction%
  \label{introduction}%
}

One of the first things a scientist in a data-intensive field hears about statistics is that there are two different approaches: frequentism and Bayesianism. Despite their importance, many researchers never have opportunity to learn the distinctions between them and the different practical approaches that result.

This paper seeks to synthesize the philosophical and pragmatic aspects of this debate, so that scientists who use these approaches might be better prepared to understand the tools available to them. Along the way we will explore the fundamental philosophical disagreement between frequentism and Bayesianism, explore the practical aspects of how this disagreement affects data analysis, and discuss the ways that these practices may affect the interpretation of scientific results.

This paper is written for scientists who have picked up some statistical knowledge along the way, but who may not fully appreciate the philosophical differences between frequentist and Bayesian approaches and the effect these differences have on both the computation and interpretation of statistical results. Because this passing statistics knowledge generally leans toward frequentist principles, this paper will go into more depth on the details of Bayesian rather than frequentist approaches. Still, it is not meant to be a full introduction to either class of methods. In particular, concepts such as the likelihood are assumed rather than derived, and many advanced Bayesian and frequentist diagnostic tests are left out in favor of illustrating the most fundamental aspects of the approaches. For a more complete treatment, see, e.g. \cite{Wasserman2004} or \cite{Gelman2004}.

\subsection{The Disagreement: The Definition of Probability%
  \label{the-disagreement-the-definition-of-probability}%
}

Fundamentally, the disagreement between frequentists and Bayesians concerns the definition of probability.

For frequentists, probability only has meaning in terms of \textbf{a limiting case of repeated measurements}. That is, if an astronomer measures the photon flux $F$ from a given non-variable star, then measures it again, then again, and so on, each time the result will be slightly different due to the statistical error of the measuring device. In the limit of many measurements, the \emph{frequency} of any given value indicates the probability of measuring that value.  For frequentists, \textbf{probabilities are fundamentally related to frequencies of events}. This means, for example, that in a strict frequentist view, it is meaningless to talk about the probability of the \emph{true} flux of the star: the true flux is, by definition, a single fixed value, and to talk about an extended frequency distribution for a fixed value is nonsense.

For Bayesians, the concept of probability is extended to cover \textbf{degrees of certainty about statements}. A Bayesian might claim to know the flux $F$ of a star with some probability $P(F)$: that probability can certainly be estimated from frequencies in the limit of a large number of repeated experiments, but this is not fundamental. The probability is a statement of the researcher's knowledge of what the true flux is. For Bayesians, \textbf{probabilities are fundamentally related to their own knowledge about an event}. This means, for example, that in a Bayesian view, we can meaningfully talk about the probability that the \emph{true} flux of a star lies in a given range.  That probability codifies our knowledge of the value based on prior information and available data.

The surprising thing is that this arguably subtle difference in philosophy can lead, in practice, to vastly different approaches to the statistical analysis of data.  Below we will explore a few examples chosen to illustrate the differences in approach, along with associated Python code to demonstrate the practical aspects of the frequentist and Bayesian approaches.

\subsection{A Simple Example: Photon Flux Measurements%
  \label{a-simple-example-photon-flux-measurements}%
}

First we will compare the frequentist and Bayesian approaches to the solution of an extremely simple problem. Imagine that we point a telescope to the sky, and observe the light coming from a single star. For simplicity, we will assume that the star's true photon flux is constant with time, i.e. that is it has a fixed value $F$; we will also ignore effects like sky background systematic errors. We will assume that a series of $N$ measurements are performed, where the $i^{\rm th}$ measurement reports the observed flux $F_i$ and error $e_i$.\DUfootnotemark{id5}{note-about-errors}{2}  The question is, given this set of measurements $D = \{F_i,e_i\}$, what is our best estimate of the true flux $F$?%
\DUfootnotetext{note-about-errors}{id5}{2}{
We will make the reasonable assumption of normally-distributed measurement errors. In a Frequentist perspective, $e_i$ is the standard deviation of the results of the single measurement event in the limit of (imaginary) repetitions of \emph{that event}. In the Bayesian perspective, $e_i$ describes the probability distribution which quantifies our knowledge of $F$ given the measured value $F_i$.}

First we will use Python to generate some toy data to demonstrate the two approaches to the problem. We will draw 50 samples $F_i$ with a mean of 1000 (in arbitrary units) and a (known) error $e_i$:\vspace{1mm}
\begin{Verbatim}[commandchars=\\\{\},fontsize=\footnotesize]
\PY{o}{\PYZgt{}\PYZgt{}}\PY{o}{\PYZgt{}} \PY{n}{np}\PY{o}{.}\PY{n}{random}\PY{o}{.}\PY{n}{seed}\PY{p}{(}\PY{l+m+mi}{2}\PY{p}{)}  \PY{c}{\PYZsh{} for reproducibility}
\PY{o}{\PYZgt{}\PYZgt{}}\PY{o}{\PYZgt{}} \PY{n}{e} \PY{o}{=} \PY{n}{np}\PY{o}{.}\PY{n}{random}\PY{o}{.}\PY{n}{normal}\PY{p}{(}\PY{l+m+mi}{30}\PY{p}{,} \PY{l+m+mi}{3}\PY{p}{,} \PY{l+m+mi}{50}\PY{p}{)}
\PY{o}{\PYZgt{}\PYZgt{}}\PY{o}{\PYZgt{}} \PY{n}{F} \PY{o}{=} \PY{n}{np}\PY{o}{.}\PY{n}{random}\PY{o}{.}\PY{n}{normal}\PY{p}{(}\PY{l+m+mi}{1000}\PY{p}{,} \PY{n}{e}\PY{p}{)}
\end{Verbatim}
\vspace{1mm}
In this toy example we already know the true flux $F$, but the question is this: given our measurements and errors, what is our best point estimate of the true flux? Let's look at a frequentist and a Bayesian approach to solving this.

\subsubsection{Frequentist Approach to Flux Measurement%
  \label{frequentist-approach-to-flux-measurement}%
}

We will start with the classical frequentist maximum likelihood approach. Given a single observation $D_i = (F_i, e_i)$, we can compute the probability distribution of the measurement given the true flux $F$ given our assumption of Gaussian errors:\begin{equation*}
P(D_i|F) = \left(2\pi e_i^2\right)^{-1/2} \exp{\left(\frac{-(F_i - F)^2}{2 e_i^2}\right)}.
\end{equation*}This should be read ``the probability of $D_i$ given $F$ equals ...''. You should recognize this as a normal distribution with mean $F$ and standard deviation $e_i$. We construct the \emph{likelihood} by computing the product of the probabilities for each data point:\begin{equation*}
\mathcal{L}(D|F) = \prod_{i=1}^N P(D_i|F)
\end{equation*}Here $D = \{D_i\}$ represents the entire set of measurements. For reasons of both analytic simplicity and numerical accuracy, it is often more convenient to instead consider the log-likelihood; combining the previous two equations gives\begin{equation*}
\log\mathcal{L}(D|F) = -\frac{1}{2} \sum_{i=1}^N \left[ \log(2\pi  e_i^2) + \frac{(F_i - F)^2}{e_i^2} \right].
\end{equation*}We would like to determine the value of $F$ which maximizes the likelihood. For this simple problem, the maximization can be computed analytically (e.g. by setting $d\log\mathcal{L}/dF|_{\hat{F}} = 0$), which results in the following point estimate of $F$:\begin{equation*}
\hat{F} = \frac{\sum w_i F_i}{\sum w_i};~~w_i = 1/e_i^2
\end{equation*}The result is a simple weighted mean of the observed values. Notice that in the case of equal errors $e_i$, the weights cancel and $\hat{F}$ is simply the mean of the observed data.

We can go further and ask what the uncertainty of our estimate is. One way this can be accomplished in the frequentist approach is to construct a Gaussian approximation to the peak likelihood; in this simple case the fit can be solved analytically to give:\begin{equation*}
\sigma_{\hat{F}} = \left(\sum_{i=1}^N w_i \right)^{-1/2}
\end{equation*}This result can be evaluated this in Python as follows:\vspace{1mm}
\begin{Verbatim}[commandchars=\\\{\},fontsize=\footnotesize]
\PY{o}{\PYZgt{}\PYZgt{}}\PY{o}{\PYZgt{}} \PY{n}{w} \PY{o}{=} \PY{l+m+mf}{1.} \PY{o}{/} \PY{n}{e} \PY{o}{*}\PY{o}{*} \PY{l+m+mi}{2}
\PY{o}{\PYZgt{}\PYZgt{}}\PY{o}{\PYZgt{}} \PY{n}{F\PYZus{}hat} \PY{o}{=} \PY{n}{np}\PY{o}{.}\PY{n}{sum}\PY{p}{(}\PY{n}{w} \PY{o}{*} \PY{n}{F}\PY{p}{)} \PY{o}{/} \PY{n}{np}\PY{o}{.}\PY{n}{sum}\PY{p}{(}\PY{n}{w}\PY{p}{)}
\PY{o}{\PYZgt{}\PYZgt{}}\PY{o}{\PYZgt{}} \PY{n}{sigma\PYZus{}F} \PY{o}{=} \PY{n}{w}\PY{o}{.}\PY{n}{sum}\PY{p}{(}\PY{p}{)} \PY{o}{*}\PY{o}{*} \PY{o}{\PYZhy{}}\PY{l+m+mf}{0.5}
\end{Verbatim}
\vspace{1mm}
For our particular data, the result is $\hat{F} = 999 \pm 4$.

\subsubsection{Bayesian Approach to Flux Measurement%
  \label{bayesian-approach-to-flux-measurement}%
}

The Bayesian approach, as you might expect, begins and ends with probabilities. The fundamental result of interest is our knowledge of the parameters in question, codified by the probability $P(F|D)$. To compute this result, we next apply Bayes' theorem, a fundamental law of probability:\begin{equation*}
P(F|D) = \frac{P(D|F)~P(F)}{P(D)}
\end{equation*}Though Bayes' theorem is where Bayesians get their name, it is important to note that it is not this theorem itself that is controversial, but the Bayesian \emph{interpretation of probability} implied by the term $P(F|D)$. While the above formulation makes sense given the Bayesian view of probability, the setup is fundamentally contrary to the frequentist philosophy, which says that probabilities have no meaning for fixed model parameters like $F$. In the Bayesian conception of probability, however, this poses no problem.

Let's take a look at each of the terms in this expression:%
\begin{itemize}

\item 

$P(F|D)$: The \textbf{posterior}, which is the probability of the model parameters given the data.
\item 

$P(D|F)$: The \textbf{likelihood}, which is proportional to the $\mathcal{L}(D|F)$ used in the frequentist approach.
\item 

$P(F)$: The \textbf{model prior}, which encodes what we knew about the model before considering the data $D$.
\item 

$P(D)$: The \textbf{model evidence}, which in practice amounts to simply a normalization term.
\end{itemize}

If we set the prior $P(F) \propto 1$ (a \emph{flat prior}), we find\begin{equation*}
P(F|D) \propto \mathcal{L}(D|F).
\end{equation*}That is, with a flat prior on $F$, the Bayesian posterior is maximized at precisely the same value as the frequentist result! So despite the philosophical differences, we see that the Bayesian and frequentist point estimates are equivalent for this simple problem.

You might notice that we glossed over one important piece here: the prior, $P(F)$, which we assumed to be flat.\DUfootnotemark{id6}{note-flat}{3} The prior allows inclusion of other information into the computation, which becomes very useful in cases where multiple measurement strategies are being combined to constrain a single model (as is the case in, e.g. cosmological parameter estimation). The necessity to specify a prior, however, is one of the more controversial pieces of Bayesian analysis.%
\DUfootnotetext{note-flat}{id6}{3}{
A flat prior is an example of an improper prior: that is, it cannot be normalized. In practice, we can remedy this by imposing some bounds on possible values: say, $0 < F < F_{tot}$, the total flux of all sources in the sky. As this normalization term also appears in the denominator of Bayes' theorem, it does not affect the posterior.}

A frequentist will point out that the prior is problematic when no true prior information is available. Though it might seem straightforward to use an \textbf{uninformative prior} like the flat prior mentioned above, there are some surprising subtleties involved.\DUfootnotemark{id7}{uninformative}{4} It turns out that in many situations, a truly uninformative prior cannot exist! Frequentists point out that the subjective choice of a prior which necessarily biases the result should have no place in scientific data analysis.

A Bayesian would counter that frequentism doesn't solve this problem, but simply skirts the question. Frequentism can often be viewed as simply a special case of the Bayesian approach for some (implicit) choice of the prior: a Bayesian would say that it's better to make this implicit choice explicit, even if the choice might include some subjectivity. Furthermore, as we will see below, the question frequentism answers is not always the question the researcher wants to ask.%
\DUfootnotetext{uninformative}{id7}{4}{\phantomsection\label{uninformative}
The flat prior in this case can be motivated by maximum entropy; see, e.g. \cite{Jeffreys1946}. Still, the use of uninformative priors like this often raises eyebrows among frequentists: there are good arguments that even ``uninformative'' priors can add information; see e.g. \cite{Evans2002}.}

\subsection{Where The Results Diverge%
  \label{where-the-results-diverge}%
}

In the simple example above, the frequentist and Bayesian approaches give basically the same result. In light of this, arguments over the use of a prior and the philosophy of probability may seem frivolous. However, while it is easy to show that the two approaches are often equivalent for simple problems, it is also true that they can diverge greatly in other situations. In practice, this divergence most often makes itself most clear in two different ways:\newcounter{listcnt0}
\begin{list}{\arabic{listcnt0}.}
{
\usecounter{listcnt0}
\setlength{\rightmargin}{\leftmargin}
}

\item 

The handling of nuisance parameters: i.e. parameters which affect the final result, but are not otherwise of interest.
\item 

The different handling of uncertainty: for example, the subtle (and often overlooked) difference between frequentist confidence intervals and Bayesian credible regions.\end{list}

We will discuss examples of these below.

\subsection{Nuisance Parameters: Bayes' Billiards Game%
  \label{nuisance-parameters-bayes-billiards-game}%
}

We will start by discussing the first point: nuisance parameters. A nuisance parameter is any quantity whose value is not directly relevant to the goal of an analysis, but is nevertheless required to determine the result which is of interest. For example, we might have a situation similar to the flux measurement above, but in which the errors $e_i$ are unknown. One potential approach is to treat these errors as nuisance parameters.

Here we consider an example of nuisance parameters borrowed from \cite{Eddy2004} that, in one form or another, dates all the way back to the posthumously-published 1763 paper written by Thomas Bayes himself \cite{Bayes1763}. The setting is a gambling game in which Alice and Bob bet on the outcome of a process they can't directly observe.

Alice and Bob enter a room. Behind a curtain there is a billiard table, which they cannot see. Their friend Carol rolls a ball down the table, and marks where it lands. Once this mark is in place, Carol begins rolling new balls down the table. If the ball lands to the left of the mark, Alice gets a point; if it lands to the right of the mark, Bob gets a point.  We can assume that Carol's rolls are unbiased: that is, the balls have an equal chance of ending up anywhere on the table.  The first person to reach six points wins the game.

Here the location of the mark (determined by the first roll) can be considered a nuisance parameter: it is unknown and not of immediate interest, but it clearly must be accounted for when predicting the outcome of subsequent rolls. If this first roll settles far to the right, then subsequent rolls will favor Alice. If it settles far to the left, Bob will be favored instead.

Given this setup, we seek to answer this question: \emph{In a particular game, after eight rolls, Alice has five points and Bob has three points. What is the probability that Bob will get six points and win the game?}

Intuitively, we realize that because Alice received five of the eight points, the marker placement likely favors her. Given that she has three opportunities to get a sixth point before Bob can win, she seems to have clinched it.  But quantitatively speaking, what is the probability that Bob will persist to win?

\subsubsection{A Naïve Frequentist Approach%
  \label{a-naive-frequentist-approach}%
}

Someone following a classical frequentist approach might reason as follows:

To determine the result, we need to estimate the location of the marker. We will quantify this marker placement as a probability $p$ that any given roll lands in Alice's favor.  Because five balls out of eight fell on Alice's side of the marker, we compute the maximum likelihood estimate of $p$, given by:\begin{equation*}
\hat{p} = 5/8,
\end{equation*}a result follows in a straightforward manner from the binomial likelihood. Assuming this maximum likelihood probability, we can compute the probability that Bob will win, which requires him to get a point in each of the next three rolls. This is given by:\begin{equation*}
P(B) = (1 - \hat{p})^3
\end{equation*}Thus, we find that the probability of Bob winning is 0.053, or odds against Bob winning of 18 to 1.

\subsubsection{A Bayesian Approach%
  \label{a-bayesian-approach}%
}

A Bayesian approach to this problem involves \emph{marginalizing} (i.e. integrating) over the unknown $p$ so that, assuming the prior is accurate,  our result is agnostic to its actual value. In this vein, we will consider the following quantities:%
\begin{itemize}

\item 

$B$ = Bob Wins
\item 

$D$ = observed data, i.e. $D = (n_A, n_B) = (5, 3)$
\item 

$p$ = unknown probability that a ball lands on Alice's side during the current game
\end{itemize}

We want to compute $P(B|D)$; that is, the probability that Bob wins given the observation that Alice currently has five points to Bob's three. A Bayesian would recognize that this expression is a \emph{marginal probability} which can be computed by integrating over the joint distribution $P(B,p|D)$:\begin{equation*}
P(B|D) \equiv \int_{-\infty}^\infty P(B,p|D) {\mathrm d}p
\end{equation*}This identity follows from the definition of conditional probability, and the law of total probability: that is, it is a fundamental consequence of probability axioms and will always be true. Even a frequentist would recognize this; they would simply disagree with the interpretation of $P(p)$ as being a measure of uncertainty of knowledge of the parameter $p$.

To compute this result, we will manipulate the above expression for $P(B|D)$ until we can express it in terms of other quantities that we can compute.

We start by applying the definition of conditional probability to expand the term $P(B,p|D)$:\begin{equation*}
P(B|D) = \int P(B|p, D) P(p|D) dp
\end{equation*}Next we use Bayes' rule to rewrite $P(p|D)$:\begin{equation*}
P(B|D) = \int P(B|p, D) \frac{P(D|p)P(p)}{P(D)} dp
\end{equation*}Finally, using the same probability identity we started with, we can expand $P(D)$ in the denominator to find:\begin{equation*}
P(B|D) = \frac{\int P(B|p,D) P(D|p) P(p) dp}{\int P(D|p)P(p) dp}
\end{equation*}Now the desired probability is expressed in terms of three quantities that we can compute:%
\begin{itemize}

\item 

$P(B|p,D)$: This term is proportional to the frequentist likelihood we used above. In words: given a marker placement $p$ and Alice's 5 wins to Bob's 3, what is the probability that Bob will go on to six wins?  Bob needs three wins in a row, i.e. $P(B|p,D) = (1 - p) ^ 3$.
\item 

$P(D|p)$: this is another easy-to-compute term. In words: given a probability $p$, what is the likelihood of exactly 5 positive outcomes out of eight trials? The answer comes from the Binomial distribution: $P(D|p) \propto p^5 (1-p)^3$
\item 

$P(p)$: this is our prior on the probability $p$. By the problem definition, we can assume that $p$ is evenly drawn between 0 and 1.  That is, $P(p) \propto 1$ for $0 \le p \le 1$.
\end{itemize}

Putting this all together and simplifying gives\begin{equation*}
P(B|D) = \frac{\int_0^1 (1 - p)^6 p^5 dp}{\int_0^1 (1 - p)^3 p^5 dp}.
\end{equation*}These integrals are instances of the beta function, so we can quickly evaluate the result using scipy:\vspace{1mm}
\begin{Verbatim}[commandchars=\\\{\},fontsize=\footnotesize]
\PY{o}{\PYZgt{}\PYZgt{}}\PY{o}{\PYZgt{}} \PY{k+kn}{from} \PY{n+nn}{scipy.special} \PY{k+kn}{import} \PY{n}{beta}
\PY{o}{\PYZgt{}\PYZgt{}}\PY{o}{\PYZgt{}} \PY{n}{P\PYZus{}B\PYZus{}D} \PY{o}{=} \PY{n}{beta}\PY{p}{(}\PY{l+m+mi}{6}\PY{o}{+}\PY{l+m+mi}{1}\PY{p}{,} \PY{l+m+mi}{5}\PY{o}{+}\PY{l+m+mi}{1}\PY{p}{)} \PY{o}{/} \PY{n}{beta}\PY{p}{(}\PY{l+m+mi}{3}\PY{o}{+}\PY{l+m+mi}{1}\PY{p}{,} \PY{l+m+mi}{5}\PY{o}{+}\PY{l+m+mi}{1}\PY{p}{)}
\end{Verbatim}
\vspace{1mm}
This gives $P(B|D) = 0.091$, or odds of 10 to 1 against Bob winning.

\subsubsection{Discussion%
  \label{discussion}%
}

The Bayesian approach gives odds of 10 to 1 against Bob, while the naïve frequentist approach gives odds of 18 to 1 against Bob. So which one is correct?

For a simple problem like this, we can answer this question empirically by simulating a large number of games and count the fraction of suitable games which Bob goes on to win. This can be coded in a couple dozen lines of Python (see part II of \cite{VanderPlas2014}). The result of such a simulation confirms the Bayesian result: 10 to 1 against Bob winning.

So what is the takeaway: is frequentism wrong? Not necessarily: in this case, the incorrect result is more a matter of the approach being ``naïve'' than it being ``frequentist''. The approach above does not consider how $p$ may vary. There exist frequentist methods that can address this by, e.g. applying a transformation and conditioning of the data to isolate dependence on $p$, or by performing a Bayesian-like integral over the sampling distribution of the frequentist estimator $\hat{p}$.

Another potential frequentist response is that the question itself is posed in a way that does not lend itself to the classical, frequentist approach. A frequentist might instead hope to give the answer in terms of null tests or confidence intervals: that is, they might devise a procedure to construct limits which would provably bound the correct answer in $100\times(1 - \alpha)$ percent of similar trials, for some value of $\alpha$ – say, 0.05. We will discuss the meaning of such confidence intervals below.

There is one clear common point of these two frequentist responses: both require some degree of effort and/or special expertise in classical methods; perhaps a suitable frequentist approach would be immediately obvious to an expert statistician, but is not particularly obvious to a statistical lay-person. In this sense, it could be argued that for a problem such as this (i.e. with a well-motivated prior), Bayesianism provides a more natural framework for handling nuisance parameters: by simple algebraic manipulation of a few well-known axioms of probability interpreted in a Bayesian sense, we straightforwardly arrive at the correct answer without need for other special statistical expertise.

\subsection{Confidence vs. Credibility: Jaynes' Truncated Exponential%
  \label{confidence-vs-credibility-jaynes-truncated-exponential}%
}

A second major consequence of the philosophical difference between frequentism and Bayesianism is in the handling of uncertainty, exemplified by the standard tools of each method: frequentist confidence intervals (CIs) and Bayesian credible regions (CRs). Despite their apparent similarity, the two approaches are fundamentally different. Both are statements of probability, but the probability refers to different aspects of the computed bounds. For example, when constructing a standard 95\% bound about a parameter $\theta$:%
\begin{itemize}

\item 

A Bayesian would say: ``Given our observed data, there is a 95\% probability that the true value of $\theta$ lies within the credible region''.
\item 

A frequentist would say: ``If this experiment is repeated many times, in 95\% of these cases the computed confidence interval will contain the true $\theta$.''\DUfootnotemark{id13}{wasserman-note}{5}
\end{itemize}
\DUfootnotetext{wasserman-note}{id13}{5}{
\cite{Wasserman2004}, however, notes on p. 92 that we need not consider repetitions of the same experiment; it's sufficient to consider repetitions of any correctly-performed frequentist procedure.}

Notice the subtle difference: the Bayesian makes a statement of probability about the \emph{parameter value} given a \emph{fixed credible region}. The frequentist makes a statement of probability about the \emph{confidence interval itself} given a \emph{fixed parameter value}. This distinction follows straightforwardly from the definition of probability discussed above: the Bayesian probability is a statement of degree of knowledge about a parameter; the frequentist probability is a statement of long-term limiting frequency of quantities (such as the CI) derived from the data.

This difference must necessarily affect our interpretation of results. For example, it is common in scientific literature to see it claimed that it is 95\% certain that an unknown parameter lies within a given 95\% CI, but this is not the case! This is erroneously applying the Bayesian interpretation to a frequentist construction. This frequentist oversight can perhaps be forgiven, as under most circumstances (such as the simple flux measurement example above), the Bayesian CR and frequentist CI will more-or-less overlap. But, as we will see below, this overlap cannot always be assumed, especially in the case of non-Gaussian distributions constrained by few data points. As a result, this common misinterpretation of the frequentist CI can lead to dangerously erroneous conclusions.

To demonstrate a situation in which the frequentist confidence interval and the Bayesian credibility region do not overlap, let us turn to an example given by E.T. Jaynes, a 20th century physicist who wrote extensively on statistical inference. In his words, consider a device that\begin{quotation}%
\begin{quote}

``...will operate without failure for a time $\theta$ because of a protective chemical inhibitor injected into it; but at time $\theta$ the supply of the chemical is exhausted, and failures then commence, following the exponential failure law. It is not feasible to observe the depletion of this inhibitor directly; one can observe only the resulting failures. From data on actual failure times, estimate the time $\theta$ of guaranteed safe operation...'' \cite{Jaynes1976}
\end{quote}
\end{quotation}

Essentially, we have data $D$ drawn from the model:\begin{equation*}
P(x|\theta) = \left\{
\begin{array}{lll}
\exp(\theta - x) &,& x > \theta\\
0                &,& x < \theta
\end{array}
\right\}
\end{equation*}where $p(x|\theta)$ gives the probability of failure at time $x$, given an inhibitor which lasts for a time $\theta$. We observe some failure times, say $D = \{10, 12, 15\}$, and ask for 95\% uncertainty bounds on the value of $\theta$.

First, let's think about what common-sense would tell us. Given the model, an event can only happen after a time $\theta$. Turning this around tells us that the upper-bound for $\theta$ must be $\min(D)$. So, for our particular data, we would immediately write $\theta \le 10$. With this in mind, let's explore how a frequentist and a Bayesian approach compare to this observation.

\subsubsection{Truncated Exponential: A Frequentist Approach%
  \label{truncated-exponential-a-frequentist-approach}%
}

In the frequentist paradigm, we'd like to compute a confidence interval on the value of $\theta$. We might start by observing that the population mean is given by\begin{equation*}
E(x) = \int_0^\infty xp(x)dx = \theta + 1.
\end{equation*}So, using the sample mean as the point estimate of $E(x)$, we have an unbiased estimator for $\theta$ given by\begin{equation*}
\hat{\theta} = \frac{1}{N} \sum_{i=1}^N x_i - 1.
\end{equation*}In the large-$N$ limit, the central limit theorem tells us that the sampling distribution is normal with standard deviation given by the standard error of the mean: $\sigma_{\hat{\theta}}^2 = 1/N$, and we can write the 95\% (i.e. $2\sigma$) confidence interval as\begin{equation*}
CI_{\rm large~N} = \left(\hat{\theta} - 2 N^{-1/2},~\hat{\theta} + 2 N^{-1/2}\right)
\end{equation*}For our particular observed data, this gives a confidence interval around our unbiased estimator of $CI(\theta) = (10.2, 12.5)$, entirely above our common-sense bound of $\theta < 10$! We might hope that this discrepancy is due to our use of the large-$N$ approximation with a paltry $N=3$ samples. A more careful treatment of the problem (See \cite{Jaynes1976} or part III of \cite{VanderPlas2014}) gives the exact confidence interval $(10.2, 12.2)$: the 95\% confidence interval entirely excludes the sensible bound $\theta < 10$!

\subsubsection{Truncated Exponential: A Bayesian Approach%
  \label{truncated-exponential-a-bayesian-approach}%
}

A Bayesian approach to the problem starts with Bayes' rule:\begin{equation*}
P(\theta|D) = \frac{P(D|\theta)P(\theta)}{P(D)}.
\end{equation*}We use the likelihood given by\begin{equation*}
P(D|\theta) \propto \prod_{i=1}^N P(x_i|\theta)
\end{equation*}and, in the absence of other information, use an uninformative flat prior on $\theta$ to find\begin{equation*}
P(\theta|D) \propto \left\{
\begin{array}{lll}
N\exp\left[N(\theta - \min(D))\right] &,& \theta < \min(D)\\
0                &,& \theta > \min(D)
\end{array}
\right\}
\end{equation*}where $\min(D)$ is the smallest value in the data $D$, which enters because of the truncation of $P(x_i|\theta)$. Because $P(\theta|D)$ increases exponentially up to the cutoff, the shortest 95\% credibility interval $(\theta_1, \theta_2)$ will be given by $\theta_2 = \min(D)$, and $\theta_1$ given by the solution to the equation\begin{equation*}
\int_{\theta_1}^{\theta_2} P(\theta|D){\rm d}\theta = f
\end{equation*}which has the solution\begin{equation*}
\theta_1 = \theta_2 + \frac{1}{N}\ln\left[1 - f(1 - e^{-N\theta_2})\right].
\end{equation*}For our particular data, the Bayesian credible region is\begin{equation*}
CR(\theta) = (9.0, 10.0)
\end{equation*}which agrees with our common-sense bound.

\subsubsection{Discussion%
  \label{id18}%
}

Why do the frequentist CI and Bayesian CR give such different results? The reason goes back to the definitions of the CI and CR, and to the fact that \emph{the two approaches are answering different questions}. The Bayesian CR answers a question about the value of $\theta$ itself (the probability that the parameter is in the fixed CR), while the frequentist CI answers a question about the procedure used to construct the CI (the probability that any potential CI will contain the fixed parameter).

Using Monte Carlo simulations, it is possible to confirm that both the above results correctly answer their respective questions (see \cite{VanderPlas2014}, III). In particular, 95\% of frequentist CIs constructed using data drawn from this model in fact contain the true $\theta$. Our particular data are simply among the unhappy 5\% which the confidence interval misses. But this makes clear the danger of misapplying the Bayesian interpretation to a CI: this particular CI is not 95\% likely to contain the true value of $\theta$; it is in fact 0\% likely!

This shows that when using frequentist methods on fixed data, we must carefully keep in mind what question frequentism is answering. Frequentism does not seek a \emph{probabilistic statement about a fixed interval} as the Bayesian approach does; it instead seeks probabilistic statements about an \emph{ensemble of constructed intervals}, with the particular computed interval just a single draw from among them. Despite this, it is common to see a 95\% confidence interval interpreted in the Bayesian sense: as a fixed interval that the parameter is expected to be found in 95\% of the time. As seen clearly here, this interpretation is flawed, and should be carefully avoided.

Though we used a correct unbiased frequentist estimator above, it should be emphasized that the unbiased estimator is not always optimal for any given problem: especially one with small $N$ and/or censored models; see, e.g. \cite{Hardy2003}. Other frequentist estimators are available: for example, if the (biased) maximum likelihood estimator were used here instead, the confidence interval would be very similar to the Bayesian credible region derived above. Regardless of the choice of frequentist estimator, however, the correct interpretation of the CI is the same: it gives probabilities concerning the \emph{recipe for constructing limits}, not for the \emph{parameter values given the observed data}. For sensible parameter constraints from a single dataset, Bayesianism may be preferred, especially if the difficulties of uninformative priors can be avoided through the use of true prior information.

\subsection{Bayesianism in Practice: Markov Chain Monte Carlo%
  \label{bayesianism-in-practice-markov-chain-monte-carlo}%
}

Though Bayesianism has some nice features in theory, in practice it can be extremely computationally intensive: while simple problems like those examined above lend themselves to relatively easy analytic integration, real-life Bayesian computations often require numerical integration of high-dimensional parameter spaces.

A turning-point in practical Bayesian computation was the development and application of sampling methods such as Markov Chain Monte Carlo (MCMC). MCMC is a class of algorithms which can efficiently characterize even high-dimensional posterior distributions through drawing of randomized samples such that the points are distributed according to the posterior. A detailed discussion of MCMC is well beyond the scope of this paper; an excellent introduction can be found in \cite{Gelman2004}. Below, we will propose a straightforward model and compare a standard frequentist approach with three MCMC implementations available in Python.

\subsection{Application: A Simple Linear Model%
  \label{application-a-simple-linear-model}%
}

As an example of a more realistic data-driven analysis, let's consider a simple three-parameter linear model which fits a straight-line to data with unknown errors. The parameters will be the the y-intercept $\alpha$, the slope $\beta$, and the (unknown) normal scatter $\sigma$ about the line.

For data $D = \{x_i, y_i\}$, the model is\begin{equation*}
\hat{y}(x_i|\alpha,\beta) = \alpha + \beta x_i,
\end{equation*}and the likelihood is the product of the Gaussian distribution for each point:\begin{equation*}
\mathcal{L}(D|\alpha,\beta,\sigma) = (2\pi\sigma^2)^{-N/2} \prod_{i=1}^N \exp\left[\frac{-[y_i - \hat{y}(x_i|\alpha, \beta)]^2}{2\sigma^2}\right].
\end{equation*}We will evaluate this model on the following data set:\vspace{1mm}
\begin{Verbatim}[commandchars=\\\{\},fontsize=\footnotesize]
\PY{k+kn}{import} \PY{n+nn}{numpy} \PY{k+kn}{as} \PY{n+nn}{np}
\PY{n}{np}\PY{o}{.}\PY{n}{random}\PY{o}{.}\PY{n}{seed}\PY{p}{(}\PY{l+m+mi}{42}\PY{p}{)}  \PY{c}{\PYZsh{} for repeatability}
\PY{n}{theta\PYZus{}true} \PY{o}{=} \PY{p}{(}\PY{l+m+mi}{25}\PY{p}{,} \PY{l+m+mf}{0.5}\PY{p}{)}
\PY{n}{xdata} \PY{o}{=} \PY{l+m+mi}{100} \PY{o}{*} \PY{n}{np}\PY{o}{.}\PY{n}{random}\PY{o}{.}\PY{n}{random}\PY{p}{(}\PY{l+m+mi}{20}\PY{p}{)}
\PY{n}{ydata} \PY{o}{=} \PY{n}{theta\PYZus{}true}\PY{p}{[}\PY{l+m+mi}{0}\PY{p}{]} \PY{o}{+} \PY{n}{theta\PYZus{}true}\PY{p}{[}\PY{l+m+mi}{1}\PY{p}{]} \PY{o}{*} \PY{n}{xdata}
\PY{n}{ydata} \PY{o}{=} \PY{n}{np}\PY{o}{.}\PY{n}{random}\PY{o}{.}\PY{n}{normal}\PY{p}{(}\PY{n}{ydata}\PY{p}{,} \PY{l+m+mi}{10}\PY{p}{)} \PY{c}{\PYZsh{} add error}
\end{Verbatim}
\vspace{1mm}
Below we will consider a frequentist solution to this problem computed with the statsmodels package\DUfootnotemark{id22}{statsmodels}{6}, as well as a Bayesian solution computed with several MCMC implementations in Python: emcee\DUfootnotemark{id23}{emcee}{7}, PyMC\DUfootnotemark{id24}{pymc}{8}, and PyStan\DUfootnotemark{id25}{pystan}{9}. A full discussion of the strengths and weaknesses of the various MCMC algorithms used by the packages is out of scope for this paper, as is a full discussion of performance benchmarks for the packages. Rather, the purpose of this section is to show side-by-side examples of the Python APIs of the packages. First, though, we will consider a frequentist solution.%
\DUfootnotetext{statsmodels}{id22}{6}{\phantomsection\label{statsmodels}
statsmodels: Statistics in Python \url{http://statsmodels.sourceforge.net/}}
\DUfootnotetext{emcee}{id23}{7}{\phantomsection\label{emcee}
emcee: The MCMC Hammer \url{http://dan.iel.fm/emcee/}}
\DUfootnotetext{pymc}{id24}{8}{\phantomsection\label{pymc}
PyMC: Bayesian Inference in Python \url{http://pymc-devs.github.io/pymc/}}
\DUfootnotetext{pystan}{id25}{9}{\phantomsection\label{pystan}
PyStan: The Python Interface to Stan \url{https://pystan.readthedocs.org/}}

\subsubsection{Frequentist Solution%
  \label{frequentist-solution}%
}

A frequentist solution can be found by computing the maximum likelihood point estimate. For standard linear problems such as this, the result can be computed using efficient linear algebra. If we define the \emph{parameter vector}, $\theta = [\alpha~\beta]^T$; the \emph{response vector}, $Y = [y_1~y_2~y_3~\cdots~y_N]^T$; and the \emph{design matrix},\begin{equation*}
X = \left[
       \begin{array}{lllll}
           1 & 1 & 1 &\cdots & 1\\
           x_1 & x_2 & x_3 & \cdots & x_N
       \end{array}\right]^T,
\end{equation*}it can be shown that the maximum likelihood solution is\begin{equation*}
\hat{\theta} = (X^TX)^{-1}(X^T Y).
\end{equation*}The confidence interval around this value is an ellipse in parameter space defined by the following matrix:\begin{equation*}
\Sigma_{\hat{\theta}}
               \equiv \left[
                  \begin{array}{ll}
                     \sigma_\alpha^2 & \sigma_{\alpha\beta} \\
                      \sigma_{\alpha\beta} & \sigma_\beta^2
                  \end{array}
                \right]
               = \sigma^2 (M^TM)^{-1}.
\end{equation*}Here $\sigma$ is our unknown error term; it can be estimated based on the variance of the residuals about the fit. The off-diagonal elements of $\Sigma_{\hat{\theta}}$ are the correlated uncertainty between the estimates. In code, the computation looks like this:\vspace{1mm}
\begin{Verbatim}[commandchars=\\\{\},fontsize=\footnotesize]
\PY{o}{\PYZgt{}\PYZgt{}}\PY{o}{\PYZgt{}} \PY{n}{X} \PY{o}{=} \PY{n}{np}\PY{o}{.}\PY{n}{vstack}\PY{p}{(}\PY{p}{[}\PY{n}{np}\PY{o}{.}\PY{n}{ones\PYZus{}like}\PY{p}{(}\PY{n}{xdata}\PY{p}{)}\PY{p}{,} \PY{n}{xdata}\PY{p}{]}\PY{p}{)}\PY{o}{.}\PY{n}{T}
\PY{o}{\PYZgt{}\PYZgt{}}\PY{o}{\PYZgt{}} \PY{n}{theta\PYZus{}hat} \PY{o}{=} \PY{n}{np}\PY{o}{.}\PY{n}{linalg}\PY{o}{.}\PY{n}{solve}\PY{p}{(}\PY{n}{np}\PY{o}{.}\PY{n}{dot}\PY{p}{(}\PY{n}{X}\PY{o}{.}\PY{n}{T}\PY{p}{,} \PY{n}{X}\PY{p}{)}\PY{p}{,}
\PY{o}{.}\PY{o}{.}\PY{o}{.}                             \PY{n}{np}\PY{o}{.}\PY{n}{dot}\PY{p}{(}\PY{n}{X}\PY{o}{.}\PY{n}{T}\PY{p}{,} \PY{n}{ydata}\PY{p}{)}\PY{p}{)}
\PY{o}{\PYZgt{}\PYZgt{}}\PY{o}{\PYZgt{}} \PY{n}{y\PYZus{}hat} \PY{o}{=} \PY{n}{np}\PY{o}{.}\PY{n}{dot}\PY{p}{(}\PY{n}{X}\PY{p}{,} \PY{n}{theta\PYZus{}hat}\PY{p}{)}
\PY{o}{\PYZgt{}\PYZgt{}}\PY{o}{\PYZgt{}} \PY{n}{sigma\PYZus{}hat} \PY{o}{=} \PY{n}{np}\PY{o}{.}\PY{n}{std}\PY{p}{(}\PY{n}{ydata} \PY{o}{\PYZhy{}} \PY{n}{y\PYZus{}hat}\PY{p}{)}
\PY{o}{\PYZgt{}\PYZgt{}}\PY{o}{\PYZgt{}} \PY{n}{Sigma} \PY{o}{=} \PY{n}{sigma\PYZus{}hat} \PY{o}{*}\PY{o}{*} \PY{l+m+mi}{2} \PY{o}{*}\PYZbs{}
\PY{o}{.}\PY{o}{.}\PY{o}{.}              \PY{n}{np}\PY{o}{.}\PY{n}{linalg}\PY{o}{.}\PY{n}{inv}\PY{p}{(}\PY{n}{np}\PY{o}{.}\PY{n}{dot}\PY{p}{(}\PY{n}{X}\PY{o}{.}\PY{n}{T}\PY{p}{,} \PY{n}{X}\PY{p}{)}\PY{p}{)}
\end{Verbatim}
\vspace{1mm}
The $1\sigma$ and $2\sigma$ results are shown by the black ellipses in Figure \DUrole{ref}{fig1}.

In practice, the frequentist approach often relies on many more statistal diagnostics beyond the maximum likelihood and confidence interval. These can be computed quickly using convenience routines built-in to the \texttt{statsmodels} package \cite{Seabold2010}. For this problem, it can be used as follows:\vspace{1mm}
\begin{Verbatim}[commandchars=\\\{\},fontsize=\footnotesize]
\PY{o}{\PYZgt{}\PYZgt{}}\PY{o}{\PYZgt{}} \PY{k+kn}{import} \PY{n+nn}{statsmodels.api} \PY{k+kn}{as} \PY{n+nn}{sm}  \PY{c}{\PYZsh{} version 0.5}
\PY{o}{\PYZgt{}\PYZgt{}}\PY{o}{\PYZgt{}} \PY{n}{X} \PY{o}{=} \PY{n}{sm}\PY{o}{.}\PY{n}{add\PYZus{}constant}\PY{p}{(}\PY{n}{xdata}\PY{p}{)}
\PY{o}{\PYZgt{}\PYZgt{}}\PY{o}{\PYZgt{}} \PY{n}{result} \PY{o}{=} \PY{n}{sm}\PY{o}{.}\PY{n}{OLS}\PY{p}{(}\PY{n}{ydata}\PY{p}{,} \PY{n}{X}\PY{p}{)}\PY{o}{.}\PY{n}{fit}\PY{p}{(}\PY{p}{)}
\PY{o}{\PYZgt{}\PYZgt{}}\PY{o}{\PYZgt{}} \PY{n}{sigma\PYZus{}hat} \PY{o}{=} \PY{n}{result}\PY{o}{.}\PY{n}{params}
\PY{o}{\PYZgt{}\PYZgt{}}\PY{o}{\PYZgt{}} \PY{n}{Sigma} \PY{o}{=} \PY{n}{result}\PY{o}{.}\PY{n}{cov\PYZus{}params}\PY{p}{(}\PY{p}{)}
\PY{o}{\PYZgt{}\PYZgt{}}\PY{o}{\PYZgt{}} \PY{k}{print}\PY{p}{(}\PY{n}{result}\PY{o}{.}\PY{n}{summary2}\PY{p}{(}\PY{p}{)}\PY{p}{)}
\end{Verbatim}
\vspace{1mm}
\vspace{1mm}
\begin{Verbatim}[commandchars=\\\{\},fontsize=\footnotesize]
====================================================
Model:              OLS  AIC:                147.773
Dependent Variable: y    BIC:                149.765
No. Observations:   20   Log\PYZhy{}Likelihood:     \PYZhy{}71.887
Df Model:           1    F\PYZhy{}statistic:        41.97
Df Residuals:       18   Prob (F\PYZhy{}statistic): 4.3e\PYZhy{}06
R\PYZhy{}squared:          0.70 Scale:              86.157
Adj. R\PYZhy{}squared:     0.68
\PYZhy{}\PYZhy{}\PYZhy{}\PYZhy{}\PYZhy{}\PYZhy{}\PYZhy{}\PYZhy{}\PYZhy{}\PYZhy{}\PYZhy{}\PYZhy{}\PYZhy{}\PYZhy{}\PYZhy{}\PYZhy{}\PYZhy{}\PYZhy{}\PYZhy{}\PYZhy{}\PYZhy{}\PYZhy{}\PYZhy{}\PYZhy{}\PYZhy{}\PYZhy{}\PYZhy{}\PYZhy{}\PYZhy{}\PYZhy{}\PYZhy{}\PYZhy{}\PYZhy{}\PYZhy{}\PYZhy{}\PYZhy{}\PYZhy{}\PYZhy{}\PYZhy{}\PYZhy{}\PYZhy{}\PYZhy{}\PYZhy{}\PYZhy{}\PYZhy{}\PYZhy{}\PYZhy{}\PYZhy{}\PYZhy{}\PYZhy{}\PYZhy{}\PYZhy{}
         Coef.  Std.Err.  t    P\PYZgt{}|t|  [0.025  0.975]
\PYZhy{}\PYZhy{}\PYZhy{}\PYZhy{}\PYZhy{}\PYZhy{}\PYZhy{}\PYZhy{}\PYZhy{}\PYZhy{}\PYZhy{}\PYZhy{}\PYZhy{}\PYZhy{}\PYZhy{}\PYZhy{}\PYZhy{}\PYZhy{}\PYZhy{}\PYZhy{}\PYZhy{}\PYZhy{}\PYZhy{}\PYZhy{}\PYZhy{}\PYZhy{}\PYZhy{}\PYZhy{}\PYZhy{}\PYZhy{}\PYZhy{}\PYZhy{}\PYZhy{}\PYZhy{}\PYZhy{}\PYZhy{}\PYZhy{}\PYZhy{}\PYZhy{}\PYZhy{}\PYZhy{}\PYZhy{}\PYZhy{}\PYZhy{}\PYZhy{}\PYZhy{}\PYZhy{}\PYZhy{}\PYZhy{}\PYZhy{}\PYZhy{}\PYZhy{}
const   24.6361  3.7871 6.5053 0.0000 16.6797 32.592
x1       0.4483  0.0692 6.4782 0.0000  0.3029  0.593
\PYZhy{}\PYZhy{}\PYZhy{}\PYZhy{}\PYZhy{}\PYZhy{}\PYZhy{}\PYZhy{}\PYZhy{}\PYZhy{}\PYZhy{}\PYZhy{}\PYZhy{}\PYZhy{}\PYZhy{}\PYZhy{}\PYZhy{}\PYZhy{}\PYZhy{}\PYZhy{}\PYZhy{}\PYZhy{}\PYZhy{}\PYZhy{}\PYZhy{}\PYZhy{}\PYZhy{}\PYZhy{}\PYZhy{}\PYZhy{}\PYZhy{}\PYZhy{}\PYZhy{}\PYZhy{}\PYZhy{}\PYZhy{}\PYZhy{}\PYZhy{}\PYZhy{}\PYZhy{}\PYZhy{}\PYZhy{}\PYZhy{}\PYZhy{}\PYZhy{}\PYZhy{}\PYZhy{}\PYZhy{}\PYZhy{}\PYZhy{}\PYZhy{}\PYZhy{}
Omnibus:          1.996    Durbin\PYZhy{}Watson:       2.75
Prob(Omnibus):    0.369    Jarque\PYZhy{}Bera (JB):    1.63
Skew:             0.651    Prob(JB):            0.44
Kurtosis:         2.486    Condition No.:       100
====================================================
\end{Verbatim}
\vspace{1mm}
The summary output includes many advanced statistics which we don't have space to fully discuss here. For a trained practitioner these diagnostics are very useful for evaluating and comparing fits, especially for more complicated models; see \cite{Wasserman2004} and the statsmodels project documentation for more details.

\subsubsection{Bayesian Solution: Overview%
  \label{bayesian-solution-overview}%
}

The Bayesian result is encapsulated in the posterior, which is proportional to the product of the likelihood and the prior; in this case we must be aware that a flat prior is not uninformative. Because of the nature of the slope, a flat prior leads to a much higher probability for steeper slopes. One might imagine addressing this by transforming variables, e.g. using a flat prior on the angle the line makes with the x-axis rather than the slope. It turns out that the appropriate change of variables can be determined much more rigorously by following arguments first developed by \cite{Jeffreys1946}.

Our model is given by $y = \alpha + \beta x$ with probability element $P(\alpha, \beta)d\alpha d\beta$. By symmetry, we could just as well have written $x = \alpha^\prime + \beta^\prime y$ with probability element $Q(\alpha^\prime, \beta^\prime)d\alpha^\prime d\beta^\prime$. It then follows that $(\alpha^\prime, \beta^\prime) = (-\beta^{-1}\alpha, \beta^{-1})$. Computing the determinant of the Jacobian of this transformation, we can then show that $Q(\alpha^\prime, \beta^\prime) = \beta^3 P(\alpha, \beta)$. The symmetry of the problem requires equivalence of $P$ and $Q$, or $\beta^3 P(\alpha,\beta) = P(-\beta^{-1}\alpha, \beta^{-1})$, which is a functional equation satisfied by\begin{equation*}
P(\alpha, \beta) \propto (1 + \beta^2)^{-3/2}.
\end{equation*}This turns out to be equivalent to choosing flat priors on the alternate variables $(\theta, \alpha_\perp) = (\tan^{-1}\beta, \alpha\cos\theta)$.

Through similar arguments based on the invariance of $\sigma$ under a change of units, we can show that\begin{equation*}
P(\sigma) \propto 1/\sigma,
\end{equation*}which is most commonly known a the \emph{Jeffreys Prior} for scale factors after \cite{Jeffreys1946}, and is equivalent to flat prior on $\log\sigma$. Putting these together, we find the following uninformative prior for our linear regression problem:\begin{equation*}
P(\alpha,\beta,\sigma) \propto \frac{1}{\sigma}(1 + \beta^2)^{-3/2}.
\end{equation*}With this prior and the above likelihood, we are prepared to numerically evaluate the posterior via MCMC.

\subsubsection{Solution with emcee%
  \label{solution-with-emcee}%
}

The emcee package \cite{ForemanMackey2013} is a lightweight pure-Python package which implements Affine Invariant Ensemble MCMC \cite{Goodman2010}, a sophisticated version of MCMC sampling. To use \texttt{emcee}, all that is required is to define a Python function representing the logarithm of the posterior. For clarity, we will factor this definition into two functions, the log-prior and the log-likelihood:\vspace{1mm}
\begin{Verbatim}[commandchars=\\\{\},fontsize=\footnotesize]
\PY{k+kn}{import} \PY{n+nn}{emcee}  \PY{c}{\PYZsh{} version 2.0}

\PY{k}{def} \PY{n+nf}{log\PYZus{}prior}\PY{p}{(}\PY{n}{theta}\PY{p}{)}\PY{p}{:}
    \PY{n}{alpha}\PY{p}{,} \PY{n}{beta}\PY{p}{,} \PY{n}{sigma} \PY{o}{=} \PY{n}{theta}
    \PY{k}{if} \PY{n}{sigma} \PY{o}{\PYZlt{}} \PY{l+m+mi}{0}\PY{p}{:}
        \PY{k}{return} \PY{o}{\PYZhy{}}\PY{n}{np}\PY{o}{.}\PY{n}{inf}  \PY{c}{\PYZsh{} log(0)}
    \PY{k}{else}\PY{p}{:}
        \PY{k}{return} \PY{p}{(}\PY{o}{\PYZhy{}}\PY{l+m+mf}{1.5} \PY{o}{*} \PY{n}{np}\PY{o}{.}\PY{n}{log}\PY{p}{(}\PY{l+m+mi}{1} \PY{o}{+} \PY{n}{beta}\PY{o}{*}\PY{o}{*}\PY{l+m+mi}{2}\PY{p}{)}
                \PY{o}{\PYZhy{}} \PY{n}{np}\PY{o}{.}\PY{n}{log}\PY{p}{(}\PY{n}{sigma}\PY{p}{)}\PY{p}{)}

\PY{k}{def} \PY{n+nf}{log\PYZus{}like}\PY{p}{(}\PY{n}{theta}\PY{p}{,} \PY{n}{x}\PY{p}{,} \PY{n}{y}\PY{p}{)}\PY{p}{:}
   \PY{n}{alpha}\PY{p}{,} \PY{n}{beta}\PY{p}{,} \PY{n}{sigma} \PY{o}{=} \PY{n}{theta}
   \PY{n}{y\PYZus{}model} \PY{o}{=} \PY{n}{alpha} \PY{o}{+} \PY{n}{beta} \PY{o}{*} \PY{n}{x}
   \PY{k}{return} \PY{o}{\PYZhy{}}\PY{l+m+mf}{0.5} \PY{o}{*} \PY{n}{np}\PY{o}{.}\PY{n}{sum}\PY{p}{(}\PY{n}{np}\PY{o}{.}\PY{n}{log}\PY{p}{(}\PY{l+m+mi}{2}\PY{o}{*}\PY{n}{np}\PY{o}{.}\PY{n}{pi}\PY{o}{*}\PY{n}{sigma}\PY{o}{*}\PY{o}{*}\PY{l+m+mi}{2}\PY{p}{)} \PY{o}{+}
                        \PY{p}{(}\PY{n}{y}\PY{o}{\PYZhy{}}\PY{n}{y\PYZus{}model}\PY{p}{)}\PY{o}{*}\PY{o}{*}\PY{l+m+mi}{2} \PY{o}{/} \PY{n}{sigma}\PY{o}{*}\PY{o}{*}\PY{l+m+mi}{2}\PY{p}{)}

\PY{k}{def} \PY{n+nf}{log\PYZus{}posterior}\PY{p}{(}\PY{n}{theta}\PY{p}{,} \PY{n}{x}\PY{p}{,} \PY{n}{y}\PY{p}{)}\PY{p}{:}
    \PY{k}{return} \PY{n}{log\PYZus{}prior}\PY{p}{(}\PY{n}{theta}\PY{p}{)} \PY{o}{+} \PY{n}{log\PYZus{}like}\PY{p}{(}\PY{n}{theta}\PY{p}{,}\PY{n}{x}\PY{p}{,}\PY{n}{y}\PY{p}{)}
\end{Verbatim}
\vspace{1mm}
Next we set up the computation. \texttt{emcee} combines multiple interacting ``walkers'', each of which results in its own Markov chain. We will also specify a burn-in period, to allow the chains to stabilize prior to drawing our final traces:\vspace{1mm}
\begin{Verbatim}[commandchars=\\\{\},fontsize=\footnotesize]
\PY{n}{ndim} \PY{o}{=} \PY{l+m+mi}{3}  \PY{c}{\PYZsh{} number of parameters in the model}
\PY{n}{nwalkers} \PY{o}{=} \PY{l+m+mi}{50}  \PY{c}{\PYZsh{} number of MCMC walkers}
\PY{n}{nburn} \PY{o}{=} \PY{l+m+mi}{1000}  \PY{c}{\PYZsh{} \PYZdq{}burn\PYZhy{}in\PYZdq{} to stabilize chains}
\PY{n}{nsteps} \PY{o}{=} \PY{l+m+mi}{2000}  \PY{c}{\PYZsh{} number of MCMC steps to take}
\PY{n}{starting\PYZus{}guesses} \PY{o}{=} \PY{n}{np}\PY{o}{.}\PY{n}{random}\PY{o}{.}\PY{n}{rand}\PY{p}{(}\PY{n}{nwalkers}\PY{p}{,} \PY{n}{ndim}\PY{p}{)}
\end{Verbatim}
\vspace{1mm}
Now we call the sampler and extract the trace:\vspace{1mm}
\begin{Verbatim}[commandchars=\\\{\},fontsize=\footnotesize]
\PY{n}{sampler} \PY{o}{=} \PY{n}{emcee}\PY{o}{.}\PY{n}{EnsembleSampler}\PY{p}{(}\PY{n}{nwalkers}\PY{p}{,} \PY{n}{ndim}\PY{p}{,}
                                \PY{n}{log\PYZus{}posterior}\PY{p}{,}
                                \PY{n}{args}\PY{o}{=}\PY{p}{[}\PY{n}{xdata}\PY{p}{,}\PY{n}{ydata}\PY{p}{]}\PY{p}{)}
\PY{n}{sampler}\PY{o}{.}\PY{n}{run\PYZus{}mcmc}\PY{p}{(}\PY{n}{starting\PYZus{}guesses}\PY{p}{,} \PY{n}{nsteps}\PY{p}{)}

\PY{c}{\PYZsh{} chain is of shape (nwalkers, nsteps, ndim):}
\PY{c}{\PYZsh{} discard burn\PYZhy{}in points and reshape:}
\PY{n}{trace} \PY{o}{=} \PY{n}{sampler}\PY{o}{.}\PY{n}{chain}\PY{p}{[}\PY{p}{:}\PY{p}{,} \PY{n}{nburn}\PY{p}{:}\PY{p}{,} \PY{p}{:}\PY{p}{]}
\PY{n}{trace} \PY{o}{=} \PY{n}{trace}\PY{o}{.}\PY{n}{reshape}\PY{p}{(}\PY{o}{\PYZhy{}}\PY{l+m+mi}{1}\PY{p}{,} \PY{n}{ndim}\PY{p}{)}\PY{o}{.}\PY{n}{T}
\end{Verbatim}
\vspace{1mm}
The result is shown by the blue curve in Figure \DUrole{ref}{fig1}.

\subsubsection{Solution with PyMC%
  \label{solution-with-pymc}%
}

The PyMC package \cite{Patil2010} is an MCMC implementation written in Python and Fortran. It makes use of the classic Metropolis-Hastings MCMC sampler \cite{Gelman2004}, and includes many built-in features, such as support for efficient sampling of common prior distributions. Because of this, it requires more specialized boilerplate than does emcee, but the result is a very powerful tool for flexible Bayesian inference.

The example below uses PyMC version 2.3; as of this writing, there exists an early release of version 3.0, which is a complete rewrite of the package with a more streamlined API and more efficient computational backend. To use PyMC, we first we define all the variables using its classes and decorators:\vspace{1mm}
\begin{Verbatim}[commandchars=\\\{\},fontsize=\footnotesize]
\PY{k+kn}{import} \PY{n+nn}{pymc}  \PY{c}{\PYZsh{} version 2.3}

\PY{n}{alpha} \PY{o}{=} \PY{n}{pymc}\PY{o}{.}\PY{n}{Uniform}\PY{p}{(}\PY{l+s}{\PYZsq{}}\PY{l+s}{alpha}\PY{l+s}{\PYZsq{}}\PY{p}{,} \PY{o}{\PYZhy{}}\PY{l+m+mi}{100}\PY{p}{,} \PY{l+m+mi}{100}\PY{p}{)}

\PY{n+nd}{@pymc.stochastic}\PY{p}{(}\PY{n}{observed}\PY{o}{=}\PY{n+nb+bp}{False}\PY{p}{)}
\PY{k}{def} \PY{n+nf}{beta}\PY{p}{(}\PY{n}{value}\PY{o}{=}\PY{l+m+mi}{0}\PY{p}{)}\PY{p}{:}
    \PY{k}{return} \PY{o}{\PYZhy{}}\PY{l+m+mf}{1.5} \PY{o}{*} \PY{n}{np}\PY{o}{.}\PY{n}{log}\PY{p}{(}\PY{l+m+mi}{1} \PY{o}{+} \PY{n}{value}\PY{o}{*}\PY{o}{*}\PY{l+m+mi}{2}\PY{p}{)}

\PY{n+nd}{@pymc.stochastic}\PY{p}{(}\PY{n}{observed}\PY{o}{=}\PY{n+nb+bp}{False}\PY{p}{)}
\PY{k}{def} \PY{n+nf}{sigma}\PY{p}{(}\PY{n}{value}\PY{o}{=}\PY{l+m+mi}{1}\PY{p}{)}\PY{p}{:}
    \PY{k}{return} \PY{o}{\PYZhy{}}\PY{n}{np}\PY{o}{.}\PY{n}{log}\PY{p}{(}\PY{n+nb}{abs}\PY{p}{(}\PY{n}{value}\PY{p}{)}\PY{p}{)}

\PY{c}{\PYZsh{} Define the form of the model and likelihood}
\PY{n+nd}{@pymc.deterministic}
\PY{k}{def} \PY{n+nf}{y\PYZus{}model}\PY{p}{(}\PY{n}{x}\PY{o}{=}\PY{n}{xdata}\PY{p}{,} \PY{n}{alpha}\PY{o}{=}\PY{n}{alpha}\PY{p}{,} \PY{n}{beta}\PY{o}{=}\PY{n}{beta}\PY{p}{)}\PY{p}{:}
    \PY{k}{return} \PY{n}{alpha} \PY{o}{+} \PY{n}{beta} \PY{o}{*} \PY{n}{x}

\PY{n}{y} \PY{o}{=} \PY{n}{pymc}\PY{o}{.}\PY{n}{Normal}\PY{p}{(}\PY{l+s}{\PYZsq{}}\PY{l+s}{y}\PY{l+s}{\PYZsq{}}\PY{p}{,} \PY{n}{mu}\PY{o}{=}\PY{n}{y\PYZus{}model}\PY{p}{,} \PY{n}{tau}\PY{o}{=}\PY{l+m+mf}{1.}\PY{o}{/}\PY{n}{sigma}\PY{o}{*}\PY{o}{*}\PY{l+m+mi}{2}\PY{p}{,}
                \PY{n}{observed}\PY{o}{=}\PY{n+nb+bp}{True}\PY{p}{,} \PY{n}{value}\PY{o}{=}\PY{n}{ydata}\PY{p}{)}

\PY{c}{\PYZsh{} package the full model in a dictionary}
\PY{n}{model} \PY{o}{=} \PY{n+nb}{dict}\PY{p}{(}\PY{n}{alpha}\PY{o}{=}\PY{n}{alpha}\PY{p}{,} \PY{n}{beta}\PY{o}{=}\PY{n}{beta}\PY{p}{,} \PY{n}{sigma}\PY{o}{=}\PY{n}{sigma}\PY{p}{,}
             \PY{n}{y\PYZus{}model}\PY{o}{=}\PY{n}{y\PYZus{}model}\PY{p}{,} \PY{n}{y}\PY{o}{=}\PY{n}{y}\PY{p}{)}
\end{Verbatim}
\vspace{1mm}
Next we run the chain and extract the trace:\vspace{1mm}
\begin{Verbatim}[commandchars=\\\{\},fontsize=\footnotesize]
\PY{n}{S} \PY{o}{=} \PY{n}{pymc}\PY{o}{.}\PY{n}{MCMC}\PY{p}{(}\PY{n}{model}\PY{p}{)}
\PY{n}{S}\PY{o}{.}\PY{n}{sample}\PY{p}{(}\PY{n+nb}{iter}\PY{o}{=}\PY{l+m+mi}{100000}\PY{p}{,} \PY{n}{burn}\PY{o}{=}\PY{l+m+mi}{50000}\PY{p}{)}
\PY{n}{trace} \PY{o}{=} \PY{p}{[}\PY{n}{S}\PY{o}{.}\PY{n}{trace}\PY{p}{(}\PY{l+s}{\PYZsq{}}\PY{l+s}{alpha}\PY{l+s}{\PYZsq{}}\PY{p}{)}\PY{p}{[}\PY{p}{:}\PY{p}{]}\PY{p}{,} \PY{n}{S}\PY{o}{.}\PY{n}{trace}\PY{p}{(}\PY{l+s}{\PYZsq{}}\PY{l+s}{beta}\PY{l+s}{\PYZsq{}}\PY{p}{)}\PY{p}{[}\PY{p}{:}\PY{p}{]}\PY{p}{,}
         \PY{n}{S}\PY{o}{.}\PY{n}{trace}\PY{p}{(}\PY{l+s}{\PYZsq{}}\PY{l+s}{sigma}\PY{l+s}{\PYZsq{}}\PY{p}{)}\PY{p}{[}\PY{p}{:}\PY{p}{]}\PY{p}{]}
\end{Verbatim}
\vspace{1mm}
The result is shown by the red curve in Figure \DUrole{ref}{fig1}.

\subsubsection{Solution with PyStan%
  \label{solution-with-pystan}%
}

PyStan is the official Python interface to Stan, a probabilistic programming language implemented in C++ and making use of a Hamiltonian MCMC using a No U-Turn Sampler \cite{Hoffman2014}. The Stan language is specifically designed for the expression of probabilistic models; PyStan lets Stan models specified in the form of Python strings be parsed, compiled, and executed by the Stan library. Because of this, PyStan is the least ``Pythonic'' of the three frameworks:\vspace{1mm}
\begin{Verbatim}[commandchars=\\\{\},fontsize=\footnotesize]
\PY{k+kn}{import} \PY{n+nn}{pystan}  \PY{c}{\PYZsh{} version 2.2}

\PY{n}{model\PYZus{}code} \PY{o}{=} \PY{l+s}{\PYZdq{}\PYZdq{}\PYZdq{}}
\PY{l+s}{data \PYZob{}}
\PY{l+s}{    int\PYZlt{}lower=0\PYZgt{} N; // number of points}
\PY{l+s}{    real x[N]; // x values}
\PY{l+s}{    real y[N]; // y values}
\PY{l+s}{\PYZcb{}}
\PY{l+s}{parameters \PYZob{}}
\PY{l+s}{    real alpha\PYZus{}perp;}
\PY{l+s}{    real\PYZlt{}lower=\PYZhy{}pi()/2, upper=pi()/2\PYZgt{} theta;}
\PY{l+s}{    real log\PYZus{}sigma;}
\PY{l+s}{\PYZcb{}}
\PY{l+s}{transformed parameters \PYZob{}}
\PY{l+s}{    real alpha;}
\PY{l+s}{    real beta;}
\PY{l+s}{    real sigma;}
\PY{l+s}{    real ymodel[N];}
\PY{l+s}{    alpha \PYZlt{}\PYZhy{} alpha\PYZus{}perp / cos(theta);}
\PY{l+s}{    beta \PYZlt{}\PYZhy{} sin(theta);}
\PY{l+s}{    sigma \PYZlt{}\PYZhy{} exp(log\PYZus{}sigma);}
\PY{l+s}{    for (j in 1:N)}
\PY{l+s}{      ymodel[j] \PYZlt{}\PYZhy{} alpha + beta * x[j];}
\PY{l+s}{    \PYZcb{}}
\PY{l+s}{model \PYZob{}}
\PY{l+s}{    y \PYZti{} normal(ymodel, sigma);}
\PY{l+s}{\PYZcb{}}
\PY{l+s}{\PYZdq{}\PYZdq{}\PYZdq{}}

\PY{c}{\PYZsh{} perform the fit \PYZam{} extract traces}
\PY{n}{data} \PY{o}{=} \PY{p}{\PYZob{}}\PY{l+s}{\PYZsq{}}\PY{l+s}{N}\PY{l+s}{\PYZsq{}}\PY{p}{:} \PY{n+nb}{len}\PY{p}{(}\PY{n}{xdata}\PY{p}{)}\PY{p}{,} \PY{l+s}{\PYZsq{}}\PY{l+s}{x}\PY{l+s}{\PYZsq{}}\PY{p}{:} \PY{n}{xdata}\PY{p}{,} \PY{l+s}{\PYZsq{}}\PY{l+s}{y}\PY{l+s}{\PYZsq{}}\PY{p}{:} \PY{n}{ydata}\PY{p}{\PYZcb{}}
\PY{n}{fit} \PY{o}{=} \PY{n}{pystan}\PY{o}{.}\PY{n}{stan}\PY{p}{(}\PY{n}{model\PYZus{}code}\PY{o}{=}\PY{n}{model\PYZus{}code}\PY{p}{,} \PY{n}{data}\PY{o}{=}\PY{n}{data}\PY{p}{,}
                  \PY{n+nb}{iter}\PY{o}{=}\PY{l+m+mi}{25000}\PY{p}{,} \PY{n}{chains}\PY{o}{=}\PY{l+m+mi}{4}\PY{p}{)}
\PY{n}{tr} \PY{o}{=} \PY{n}{fit}\PY{o}{.}\PY{n}{extract}\PY{p}{(}\PY{p}{)}
\PY{n}{trace} \PY{o}{=} \PY{p}{[}\PY{n}{tr}\PY{p}{[}\PY{l+s}{\PYZsq{}}\PY{l+s}{alpha}\PY{l+s}{\PYZsq{}}\PY{p}{]}\PY{p}{,} \PY{n}{tr}\PY{p}{[}\PY{l+s}{\PYZsq{}}\PY{l+s}{beta}\PY{l+s}{\PYZsq{}}\PY{p}{]}\PY{p}{,} \PY{n}{tr}\PY{p}{[}\PY{l+s}{\PYZsq{}}\PY{l+s}{sigma}\PY{l+s}{\PYZsq{}}\PY{p}{]}\PY{p}{]}
\end{Verbatim}
\vspace{1mm}
The result is shown by the green curve in Figure \DUrole{ref}{fig1}.

\subsubsection{Comparison%
  \label{comparison}%
}
\begin{figure}[]\noindent\makebox[\columnwidth][c]{\includegraphics[width=\columnwidth]{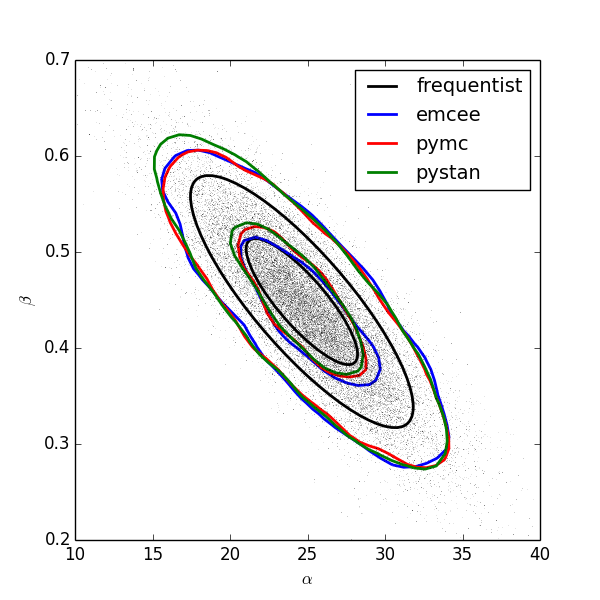}}\caption{Comparison of model fits using frequentist maximum likelihood, and Bayesian MCMC using three Python packages: emcee, PyMC, and PyStan. \DUrole{label}{fig1}}
\end{figure}

The $1\sigma$ and $2\sigma$ posterior credible regions computed with these three packages are shown beside the corresponding frequentist confidence intervals in Figure \DUrole{ref}{fig1}. The frequentist result gives slightly tighter bounds; this is primarily due to the confidence interval being computed assuming a single maximum likelihood estimate of the unknown scatter, $\sigma$ (this is analogous to the use of the single point estimate for the nuisance parameter $p$ in the billiard game, above). This interpretation can be confirmed by plotting the Bayesian posterior conditioned on the maximum likelihood estimate $\hat{\sigma}$; this gives a credible region much closer to the frequentist confidence interval.

The similarity of the three MCMC results belie the differences in algorithms used to compute them: by default, PyMC uses a Metropolis-Hastings sampler, PyStan uses a No U-Turn Sampler (NUTS), while emcee uses an affine-invariant ensemble sampler.  These approaches are known to have differing performance characteristics depending on the features of the posterior being explored. As expected for the near-Gaussian posterior used here, the three approaches give very similar results.

A main apparent difference between the packages is the Python interface. Emcee is perhaps the simplest, while PyMC requires more package-specific boilerplate code. PyStan is the most complicated, as the model specification requires directly writing a string of Stan code.

\subsection{Conclusion%
  \label{conclusion}%
}

This paper has offered a brief philosophical and practical glimpse at the differences between frequentist and Bayesian approaches to statistical analysis. These differences have their root in differing conceptions of probability: frequentists define probability as related to \emph{frequencies of repeated events}, while Bayesians define probability as a \emph{measure of uncertainty}. In practice, this means that frequentists generally quantify the properties of \emph{data-derived quantities} in light of \emph{fixed model parameters}, while Bayesians generally quantify the properties of \emph{unknown models parameters} in light of \emph{observed data}. This philosophical distinction often makes little difference in simple problems, but becomes important within more sophisticated analysis.

We first considered the case of nuisance parameters, and showed that Bayesianism offers more natural machinery to deal with nuisance parameters through \emph{marginalization}. Of course, this marginalization depends on having an accurate prior probability for the parameter being marginalized.

Next we considered the difference in the handling of uncertainty, comparing frequentist confidence intervals with Bayesian credible regions. We showed that when attempting to find a single, fixed interval bounding the true value of a parameter, the Bayesian solution answers the question that researchers most often ask. The frequentist solution can be informative; we just must be careful to correctly interpret the frequentist confidence interval.

Finally, we combined these ideas and showed several examples of the use of frequentism and Bayesianism on a more realistic linear regression problem, using several mature packages available in the Python language ecosystem. Together, these packages offer a set of tools for statistical analysis in both the frequentist and Bayesian frameworks.

So which approach is best? That is somewhat a matter of personal ideology, but also depends on the nature of the problem at hand. Frequentist approaches are often easily computed and are well-suited to truly repeatible processes and measurements, but can hit snags with small sets of data and models which depart strongly from Gaussian. Frequentist tools for these situations do exist, but often require subtle considerations and specialized expertise. Bayesian approaches require specification of a potentially subjective prior, and often involve intensive computation via MCMC. However, they are often conceptually more straightforward, and pose results in a way that is much closer to the questions a scientist wishes to answer: i.e. how do \emph{these particular data} constrain the unknowns in a certain model? When used with correct understanding of their application, both sets of  statistical tools can be used to effectively interpret of a wide variety of scientific and technical results.

\end{document}